\documentclass[%
 reprint,
nobibnotes,
 amsmath,amssymb,
 aps,
prb,
]{revtex4-2}

\usepackage{graphicx}
\usepackage{dcolumn}
\usepackage{bm}

\usepackage[dvipsnames]{xcolor}

\newcommand{\la}{\langle}
\newcommand{\ra}{\rangle}

\newcommand{\da}{\dagger}

\newcommand{\half}{\frac{1}{2}}

\newcommand{\tildeps}{\tilde{\varepsilon}}
\newcommand{\nb}{\mathbf{n}}

\newcommand{\sox}{SO$_2$}
\newcommand{\hox}{H$_2$O}
\newcommand{\dox}{D$_2$O}
\newcommand{\nox}{NO$_2$}
\newcommand{\lmax}{$L_{max}$}
\newcommand{\invcm}{cm$^{-1}$}

\begin{document}


\title{Quantum Algorithm for Calculating Molecular Vibronic Spectra}

\author{Nicolas P. D. Sawaya}
\email{nicolas.sawaya@intel.com}
\affiliation{Intel Labs, Santa Clara, California, USA}

\author{Joonsuk Huh}%
 \email{joonsukhuh@gmail.com}
 \affiliation{Department of Chemistry, Sungkyunkwan University, Republic of Korea; \\SKKU Advanced Institute of Nanotechnology (SAINT), Sungkyunkwan University, Republic of Korea}%





\date{\today}

\begin{abstract}
\textbf{Abstract.} We present a quantum algorithm for calculating the vibronic spectrum of a molecule, a useful but classically hard problem in chemistry. We show several advantages over previous quantum approaches: vibrational anharmonicity is naturally included; after measurement, some state information is preserved for further analysis; and there are potential error-related benefits. Considering four triatomic molecules, we numerically study truncation errors in the harmonic approximation. Further, in order to highlight the fact that our quantum algorithm's primary advantage over classical algorithms is in simulating anharmonic spectra, we consider the anharmonic vibronic spectrum of sulfur dioxide. In the future, our approach could aid in the design of materials with specific light-harvesting and energy transfer properties, and the general strategy is applicable to other spectral calculations in chemistry and condensed matter physics.

\end{abstract}

                              


\maketitle



\textit{TOC Graphic}

\includegraphics[width=0.3\textwidth]{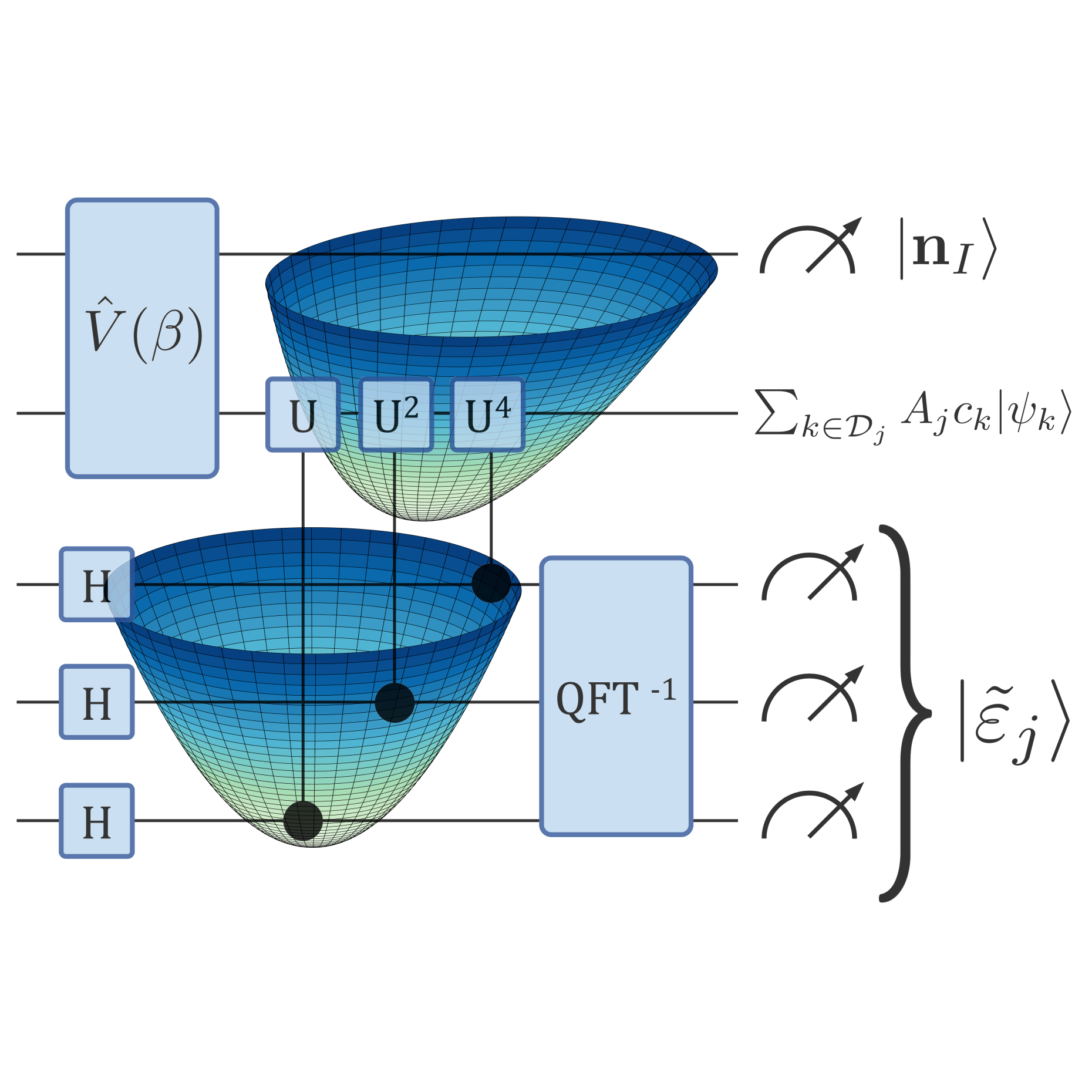}

Calculating the absorption spectrum of molecules is a common and important problem in theoretical chemistry, as it aids both the interpretation of experimental spectra and the \textit{a priori} design of molecules with particular optical properties prior to performing a costly laboratory synthesis. Further, in many molecular clusters and systems, absorption and emission spectra of molecules are required for calculating energy transfer rates \cite{may08}. 
The widespread use of mature software that solves the vibronic problem is one indication of its relevance to chemistry \cite{berger98,barone09,g16,molpro}.

Many quantum algorithms have been proposed for practical problems in chemistry, chiefly for solving the fermionic problem of determining the lowest-energy configuration of $N_e$ electrons, given the presence of a set of clamped atomic nuclei \cite{olson17,cao18,kais14,aag05,whitfield11,peruzzo14,mcclean16,babbush18,reiher17}. However, for many chemical problems of practical interest, solving the ground-state electronic structure problem is insufficient. To calculate exact vibronic spectra, for instance, an often combinatorially scaling classical algorithm must be implemented \textit{after} the electronic structure problem has been solved for many nuclear positions \cite{ruhoff00,kan08,jankowiak07,santoro07,huh11thesis,huh15}.

In this work, we propose an efficient quantum algorithm for calculating molecular vibronic spectra, within the standard quantum circuit model. A quantum algorithm to solve this problem, for implementation on a boson sampling machine \cite{huh15,huh17ft}, was previously proposed and demonstrated experimentally \cite{walmsley17,shen18}, but to our knowledge no one has previously developed an algorithm for the universal circuit model of quantum computation, nor (more importantly) one that can efficiently include vibrational anharmonic effects while calculating the full spectrum. We are also aware of unpublished work that studies the connection between quantum phase estimation and sampling problems \cite{gp18}.


Other related previous work includes quantum algorithms for calculating single Franck-Condon factors \cite{mahesh14,mcardle18} or low-lying vibrational states \cite{teplukhin18}, algorithms for simulating vibrational dynamics \cite{mcardle18,sparrow18}, and an experimental photonics implementation \cite{sparrow18} that simulated several processes related to molecular vibrations in molecules. Though these four works simulate vibrational effects, they do not address the problem of efficiently solving the \textit{full} vibronic spectrum despite the presence of an exponential number of relevant vibrational states, which is the focus of this work. Note that in this work we use the term ``classical'' solely to refer to algorithms that run on classical computers for solving the quantum problem of calculating vibronic spectra; we are \textit{not} referring to methods where nuclear degrees of freedom are approximated with Newtonian physics.



In order to calculate a vibronic spectrum, one needs to consider the transformation between two electronic potential energy surfaces (PESs). The hypersurfaces may be substantially anharmonic, making accurate classical calculations beyond a few atoms impossible \cite{luis06,huh10,meier15,petrenko17}. In order to introduce our approach, we begin by assuming that the two PESs are harmonic (\textit{i.e.} parabolic along all normal coordinates), the relationship between the two PES being defined by the Duschinsky transformation \cite{duschinsky37},

\begin{equation}\label{eq:dusch}
\vec{q'} = \textbf{S}\vec{q} + \vec{d}
\end{equation}
where $\vec q$ and $\vec{q'}$ are the vibrational normal coordinates for the initial (\textit{e.g.} ground) and final (\textit{e.g.} excited) PES, respectively, the Duschinsky matrix $\textbf{S}$ is unitary, and $\vec d$ is a displacement vector. 

\begin{figure}
  \centering
    \includegraphics[width=0.5\textwidth]{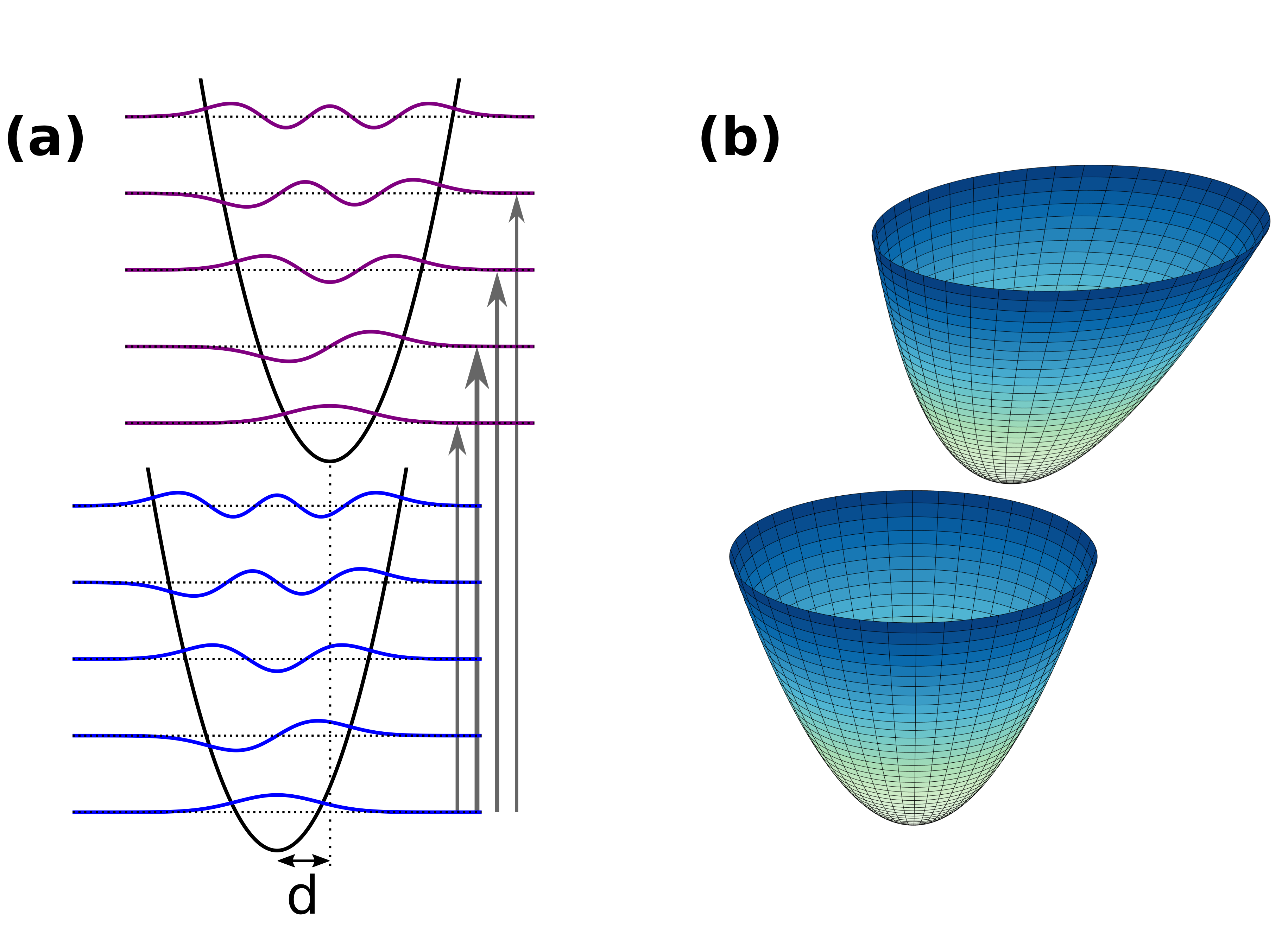}
      \caption{A schematic of the vibronic problem. \textbf{(a)} The one-dimensional case, corresponding to a diatomic molecule, for which the only vibrational degree of freedom is the distance between the two atoms. The lower and upper parabolas represent the potential energy surfaces (PESs) of the ground and first excited electronic states, respectively. The vibronic spectrum problem consists of calculating overlaps, called Franck-Condon factors (FCFs), between vibrational wavefunctions, producing a plot of intensity versus energy. In the zero-temperature case shown here, the initial state is $|0\ra$, the ground vibrational state of the ground electronic PES. The thickness of the transition arrows vary because the FCFs for those transitions differ. $d$ is the displacement vector. \textbf{(b)} The multidimensional analogue, where the normal mode coordinates of each PES are used. In the harmonic case, the relationship between the two hypersurfaces can be described by a transformation that includes displacement, squeezing, and rotation operations. For many molecules, the vibronic problem is computationally hard on a classical computer, partly because spectral contributions from exponentially many vibrational Fock states can be present.}
      \label{fig:vibronic}
\end{figure}

A vibronic spectrum calculation consists of determining the Franck-Condon profile (FCP), defined as

\begin{equation}
FCP(\omega) = \sum_{\{|f_i\ra\}}|\la0|f_i\ra|^2 \delta(\omega - \omega_{i}),
\label{eq:FCP}
\end{equation}

where $\omega$ is the transition energy, $|0\ra$ is the vibrational vacuum state of the initial PES (\textit{i.e.} of the initial electronic state), and $|f_i\ra$ is the $i$th eigenstate of the final PES with energy $\omega_i$. In practice, the function is desired to some precision $\Delta\omega$. 
Fig. \ref{fig:vibronic} gives a schematic of the vibronic problem for the (a) one-dimensional and (b) multidimensional case, where each parabola or hypersurface represent an electronic PES.




In the photonics-based vibronic boson sampling (VBS) algorithm \cite{huh15}, a change of basis known as the Doktorov transformation \cite{doktorov77} is used to transform between the two PESs, as this harmonic transformation is directly implementable in photonic circuit elements.

Instead of this direct basis change approach, our work is based on constructing a Hamiltonian that encodes the relationship between the two PESs. This provides multiple advantages, outlined below. 

We denote dimensionless position and momentum operators as $\hat q_{sk}$ and $\hat p_{sk}$ respectively, where $s$ labels the potential energy surface ($s \in \{A,B\}$ in this work) and $k$ labels the vibrational mode. These follow standard definitions $\hat q_{sk} = ( \hat a_{sk} + \hat a_{sk}^\dag )/\sqrt{2}$ and $\hat p_{sk} = ( \hat a_{sk} - \hat a_{sk}^\dag )/i\sqrt{2}$, where $a_{sk}^\dag$ and $a_{sk}$ are vibrational creation and annihilations operators. The notation $\vec {\cdot}$ denotes standard vectors as well as vectors of operators, such that e.g. $\vec{q}_A = [\hat q_{A0},...,\hat q_{AM}]^T$.


The purpose of our classical pre-processing procedure is to express the vibrational Hamiltonian for PES $B$ in terms of $\{q_{Ak},p_{Ak}\}$, by making the following transformations: $\{\vec q_A, \vec p_A\}$ $\rightarrow \{\vec Q_A, \vec P_A\}$ $\rightarrow \{\vec Q_B, \vec P_B\}$ $\rightarrow \{\vec q_B, \vec p_B\}$, where $\vec Q_s$ and $\vec P_s$ are respectively the mass-weighted position and momentum operators of PES $s$ \cite{huh11thesis}. The full transformations are



\begin{equation}\label{eq:qB}
\vec q_B = \mathbf \Omega_B \mathbf S \mathbf\Omega_A^{-1} \vec q_A + \mathbf \Omega_B \vec d 
\end{equation}

\begin{equation}\label{eq:pB}
\vec p_B = \mathbf \Omega_B^{-1} \mathbf S \mathbf\Omega_A  \vec p_A,
\end{equation}

where 
\begin{equation}
\mathbf{\Omega}_s = diag( [\omega_{s1},...,\omega_{sM}] )^{\half}
\end{equation}

and $\{\omega_{sk}\}$ are the quantum harmonic oscillator (QHO) frequencies of PES $s$. A more pedagogical explanation as well as an alternate formulation are given in the Supplemental Information. Parameter $\delta$ is often used\cite{malmqvist98,huh11thesis,huh15}, defined $\delta_{sk}=d_{sk} \sqrt{\omega_{sk}/\hbar}$.

Finally, the vibrational Hamiltonian of PES $B$ is expressed in a standard form as

\begin{equation}\label{eq:hb_qp}
    H_B = \frac{1}{2} \sum_k^M \omega_{Bk} ( q_{Bk}^2 + p_{Bk}^2),
\end{equation}





after each $q_{Bk}$ and $p_{Bk}$ has been constructed as a function of the ladder operators of PES $A$. Hence the low-level building block of our algorithm is a truncated creation operator,
\begin{equation}
\tilde a_i^\da = \sum_{l=1}^{L_{max}} \sqrt{l} | l \ra\la l-1 |
\end{equation}
where $l$ denotes a vibrational energy level and the imposed cutoff $L_{max}$ denotes the maximum level. Mappings to qubits (\textit{i.e.} integer-to-bit encodings) are discussed in the SI and errors are analyzed below. 





Though this work primarily considers the harmonic case in order to introduce our methodology, the largest quantum advantage will arise from modeling anharmonic effects. In fact, because the harmonic approximation is amenable to clever classical techniques that cannot be applied to anharmonic PESs \cite{huh11thesis}, it is expected that quantum advantage would be more easily demonstrated for the anharmonic problems than in the harmonic ones.

FCPs from anharmonicity are vastly more costly to approximate than the harmonic case using classical algorithms---for calculations that include Duschinsky and anharmonic effects, we are not aware of molecules larger than six atoms that have been accurately simulated  \cite{luis06,jankowiak07,huh10,meier15,petrenko17}. Arbitrary anharmonicity can be straightforwardly included in our quantum algorithm by adding higher-order potential energy terms to the unperturbed (e.g. Eq. \ref{eq:hb_qp}) vibrational Hamiltonian $H_0$:
\begin{equation}\label{eq:anharm}
H = H_0 + \sum_{ijk} k_{ijk} q_iq_jq_k + ... 
\end{equation}

The ease with which one includes anharmonic effects is an advantage over the VBS algorithm \cite{huh15,huh17ft}.



Now that we have outlined the required classical steps, we describe our quantum algorithm for determining the Franck-Condon profile. Unlike most quantum computational approaches to Hamiltonian simulation \cite{lloyd96,somma02,somma03,mcclean16}, which aim to find the energy of a particular quantum state, the purpose of our algorithm is to construct a full spectrum from many measurements. 

As the procedure makes use of the quantum phase estimation (QPE) algorithm \cite{kitaev97,abrams97,abrams99}, we use two quantum registers.  QPE is a quantum algorithm that calculates the eigenvalues of a superposition of states, acting on a quantum state as $\sum c_i |\psi_i\ra\otimes|0\ra$ $ \rightarrow$ $\sum c_i|\psi_i\ra\otimes|\tilde \phi_i\ra$, where $\tilde \phi_i$ and $|\psi_i\ra$ are eigenpairs with respect to an implemented operator. The first register $S$ stores a representation of the vibrational state, and the second register $E$ is used to read out the energy (strictly speaking, it outputs the phase, from which the energy is trivially obtained). 

$S$ is initialized to $|0\ra$, the ground state of $H_A$. A simple but key observation is that $|0\ra$ can be written in the eigenbasis of $H_B$, such that
\begin{equation}
|0\ra = \sum_i c_i |\psi_i\ra
\end{equation}
where $\{|\psi_i\ra\}$ are eigenstates of $H_B$ and coefficients $c_i$ are not \textit{a priori} known.

One then runs QPE using the Hamiltonian $H_B$ (\textit{i.e.} implementing $U=e^{-i\tau H_B}$ for some arbitrary value $\tau$), with register $E$ storing the eigenvalues. Many quantum algorithms have been developed for Hamiltonian simulation \cite{lloyd96,berry06,berry12,berry14,berry15a,berry15b,low17,childs18,raeisi12}, any of which can be used in conjunction with the algorithm's QPE step. Convincing numerical evidence suggests that Trotterization \cite{lloyd96,berry06} is likely to be the most viable option for early quantum devices \cite{childs18}. Computational scaling is briefly discussed in the Supplemental Information.

\begin{figure}
  \centering
    \includegraphics[width=0.5\textwidth]{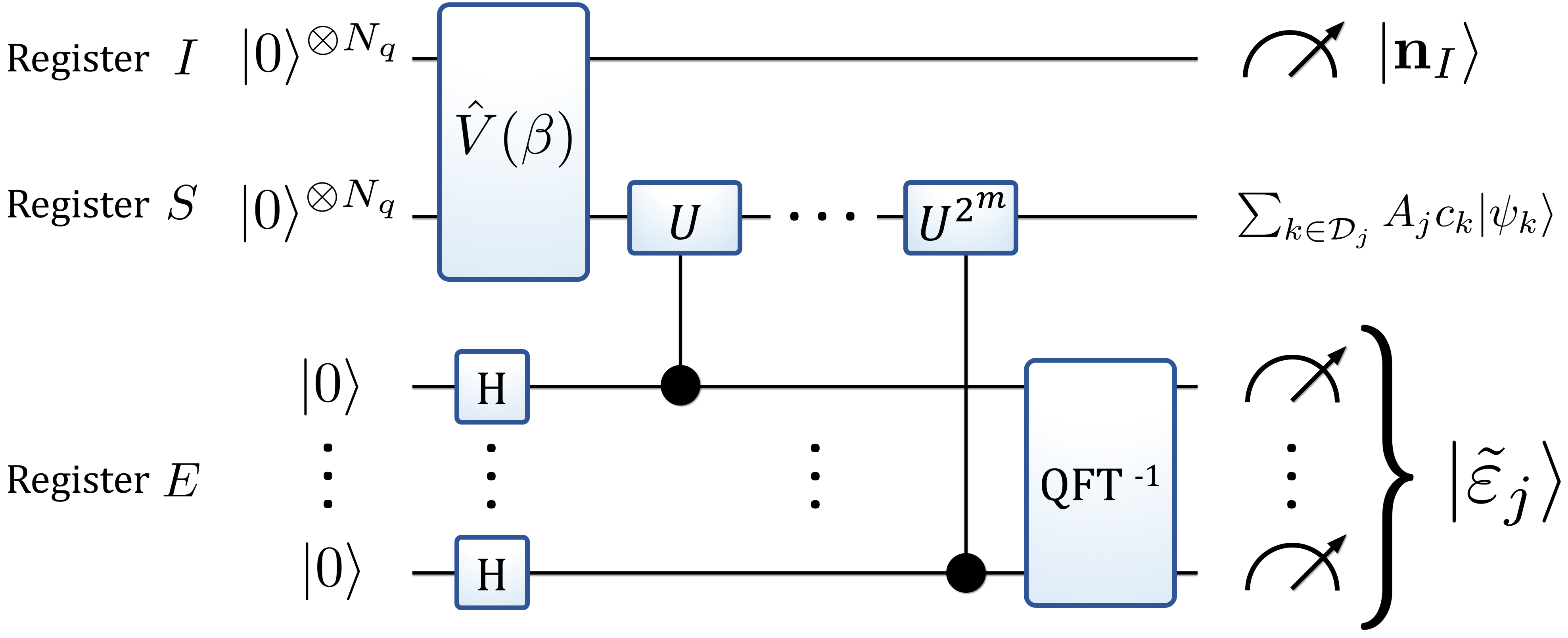}
      \caption{A quantum circuit schematic of the quantum algorithm for calculating vibronic spectra. In the zero-temperature case, only registers $S$ and $E$ are used, with gate $\hat V(\beta)$ ignored. Register $S$, which encodes the vibrational state, is initialized to $|0\ra \equiv |0\ra^{\otimes N_q}$, the vibrational ground state of the ground electronic PES. Running the quantum phase estimation (QPE) algorithm with registers $S$ and $E$ yields quantum state $\sum_i c_i |\psi_i\ra_S|\tildeps_i\ra_E$, where a key insight is that $|c_i|^2$ are proportional to the Franck-Condon Factors (FCFs) for each eigenstate $i$ of vibrational Hamiltonian $H_B$. QFT$^{-1}$ denotes the inverse quantum Fourier transform, $U$ is a unitary exponential of $H_B$, and $H$ is the Hadamard gate. A measurement on register $E$ then yields some value $\tildeps_j$, proportional to the measured phase. $\tildeps_j$ is the energy of the transition to an arbitrary precision. One then produces a histogram from many runs of the quantum circuit. Note that the quantum state $A_j\sum_{k \in \mathcal{D}_j}c_k|\psi_k\ra$ is preserved in register $S$ for further analysis. In the finite temperature case, a register $I$ (encoding the initial state) is added, and the constant-depth operation $\hat V(\beta)$ is implemented (a constant depth means that the computational cost of the operation is independent of the number of modes)). After QPE, one then measures both registers $E$ and $I$, with the contribution to the histogram being $\tildeps_j$ minus the energy of the initial Fock state $|\nb_I\ra$.}
      \label{fig:qcirc}
\end{figure}

We define $\varepsilon_i$ as the eigenenergy of $|\psi_i\ra$, and $\tildeps_i$ as its approximation, where an arbitrarily high precision can be achieved by increasing the number of qubits in register $E$. Degeneracies in $\tildeps$ will be ubiquitous, and we define the subspace of states with approximate energy $\tildeps_j$ as $\mathcal{D}_j=\{ |\psi_{j1}\ra,...,|\psi_{jK_j}\ra \}$, where $K_j$ is the degeneracy in $\tildeps_j$. Measuring register $E$ yields $\tildeps_j$ with probability $\sum_{k \in \mathcal{D}_j}|c_k|^2$. Hence---and this is the key insight---values $\tildeps_j$ are outputted with a probability \textit{exactly} in proportion to the Franck-Condon factors of Eq. \ref{eq:FCP}. The measurements then produce a histogram that yields the vibronic spectrum. The procedure is depicted in Fig. \ref{fig:qcirc}, where for the zero-temperature case one may disregard register $I$ and gate $\hat V(\beta)$. See the Supplemental Information for a step-by-step outline of the algorithm.

Note that this is a different approach from how QPE is usually used. Normally one attempts to prepare a state that is as close as possible to a desired eigenstate, whereas here we deliberately begin with a broad mix of eigenstates that corresponds to the particular spectrum we wish to calculate.

We highlight four potential benefits of this algorithm over the VBS algorithm \cite{huh15,huh17ft}. First, the quantum state in register $S$ is preserved for further analysis, while in VBS the final state is destroyed. After measurement, the state stored in $S$ is a superposition of states with energy $\tildeps_j$. From several runs of the circuit, one may estimate this stored state's overlap with another quantum state \cite{buhrman01}, estimate its expectation value with respect to an arbitrary operator, or calculate the transition energy to another PES (\textit{i.e.} simulate excited-state absorption), though methods for analyzing this preserved information are beyond the scope of this work. Our QPE-based approach has a similar benefit over the canonical quantum circuit method for calculating correlation functions \cite{somma02}, which does not provide this kind of interpretable post-measurement state.


The second potential benefit is that, as stated above, anharmonic effects are easily included in our framework. Third, accurate photon number detection for higher photon counts is a major difficulty in experimental quantum optics \cite{walmsley17,shen18}; it may be that a scaled-up universal quantum computer is built before quantum optical detectors improve satisfactorally, though this is difficult to predict. Fourth, while there are error correction methods for universal quantum computers, we do not know of such methods for boson sampling devices.


Even at room temperature, the optical spectrum of a molecule can be substantially different from its zero temperature spectrum \cite{bernath05}, necessitating methods for including finite temperature effects. These effects can be elegantly included by appending additional steps before and after the zero temperature algorithm, following previous work \cite{mann89,huh17ft}. Briefly, one appends an additional quantum register, labeled $I$, with the same size as register $S$. An operator called $\hat V(\beta)$ is applied to state $|0\ra_S|0\ra_I$, which entangles the $E$ and $I$ registers to produce a thermofield double state. The remainder of the algorithm proceeds as before, except that both $E$ and $I$ are measured and the contribution to the histogram is modified. We elaborate on this procedure in the Supplemental Information.




\begin{figure}
  \centering
    \includegraphics[width=0.5\textwidth]{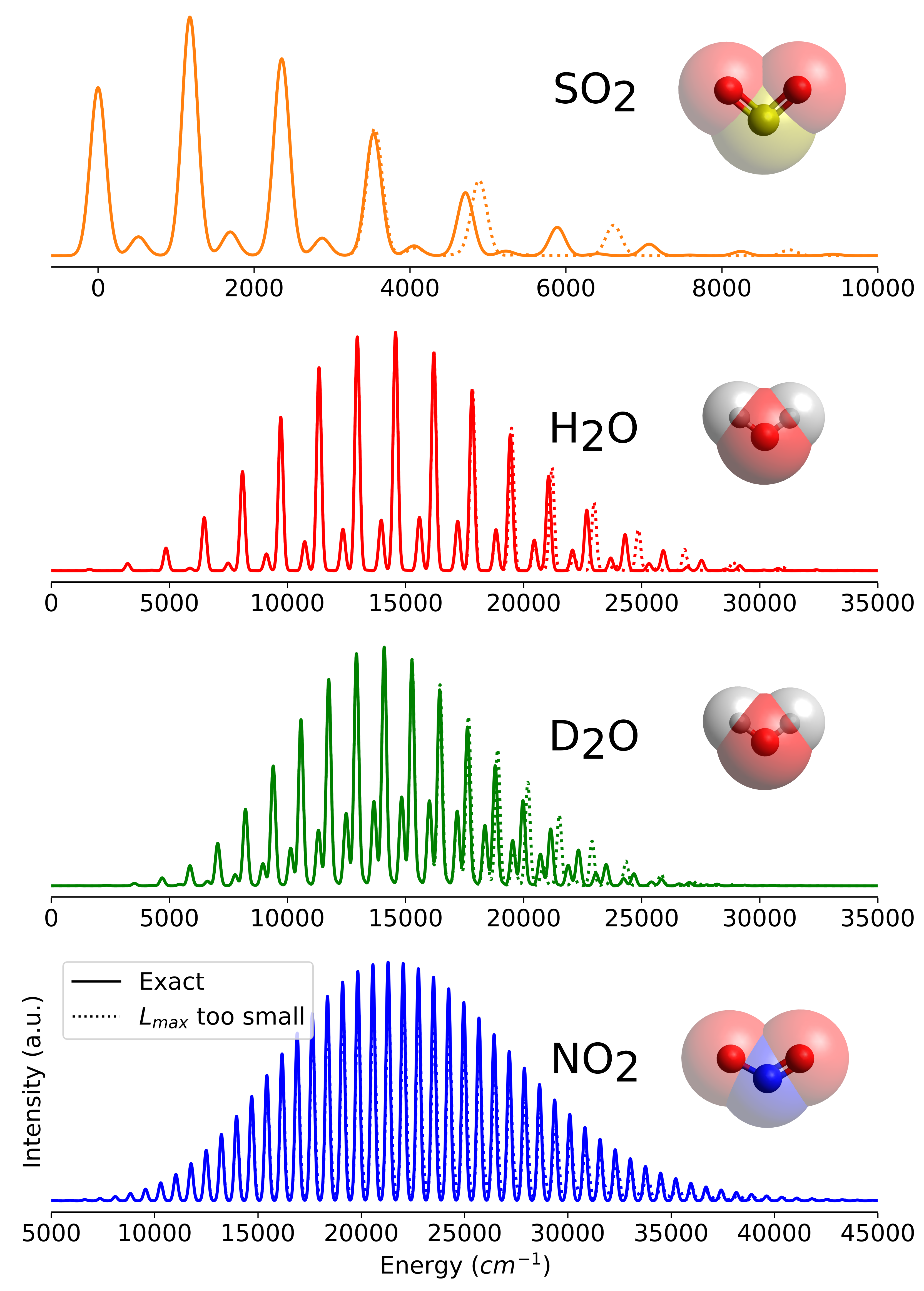}
      \caption{Theoretically exact (solid) and approximate (dotted) vibronic spectra at zero temperature under the harmonic approximation. The transitions considered are SO$_2^-$ $\rightarrow $SO$_2+$e$^-$ \cite{lee09}; H$_2$O(D$_2$O)$ \rightarrow $H$_2$O$^+$(D$_2$O$^+$)+e$^-$ \cite{chang08}; and \nox($^2A_1$) $\rightarrow  $ \nox($^2B_2$) \cite{ruhoff94} (see Supplemental Information for explicit physical parameters). After histogram construction, each peak was broadened by a Gaussian of arbitrary width 100 \invcm. Inaccuracies in the approximate spectra are due to an insufficiently large cutoff \lmax~when representing the larger-$\delta$ vibrational mode, where \lmax~is the highest energy level in the truncated ladder operator used to represent the mode. The approximate spectra are included in order to show the main qualitative effect of the truncation error, namely that lower energy peaks converge rapidly, while higher energy peaks are blue-shifted. Insight into this type of error is valuable because such truncation errors are not present in standard classical vibronic algorithms. In the approximate data, the L$_1$ errors and cutoffs \lmax~ are \{0.208, 0.231, 0.228, 0.241\} and \{10, 45, 57, 61\} respectively for \sox, \hox, \dox, and \nox.}
      \label{fig:4vibspec}
\end{figure}

\begin{figure}
  \centering
    \includegraphics[width=0.5\textwidth]{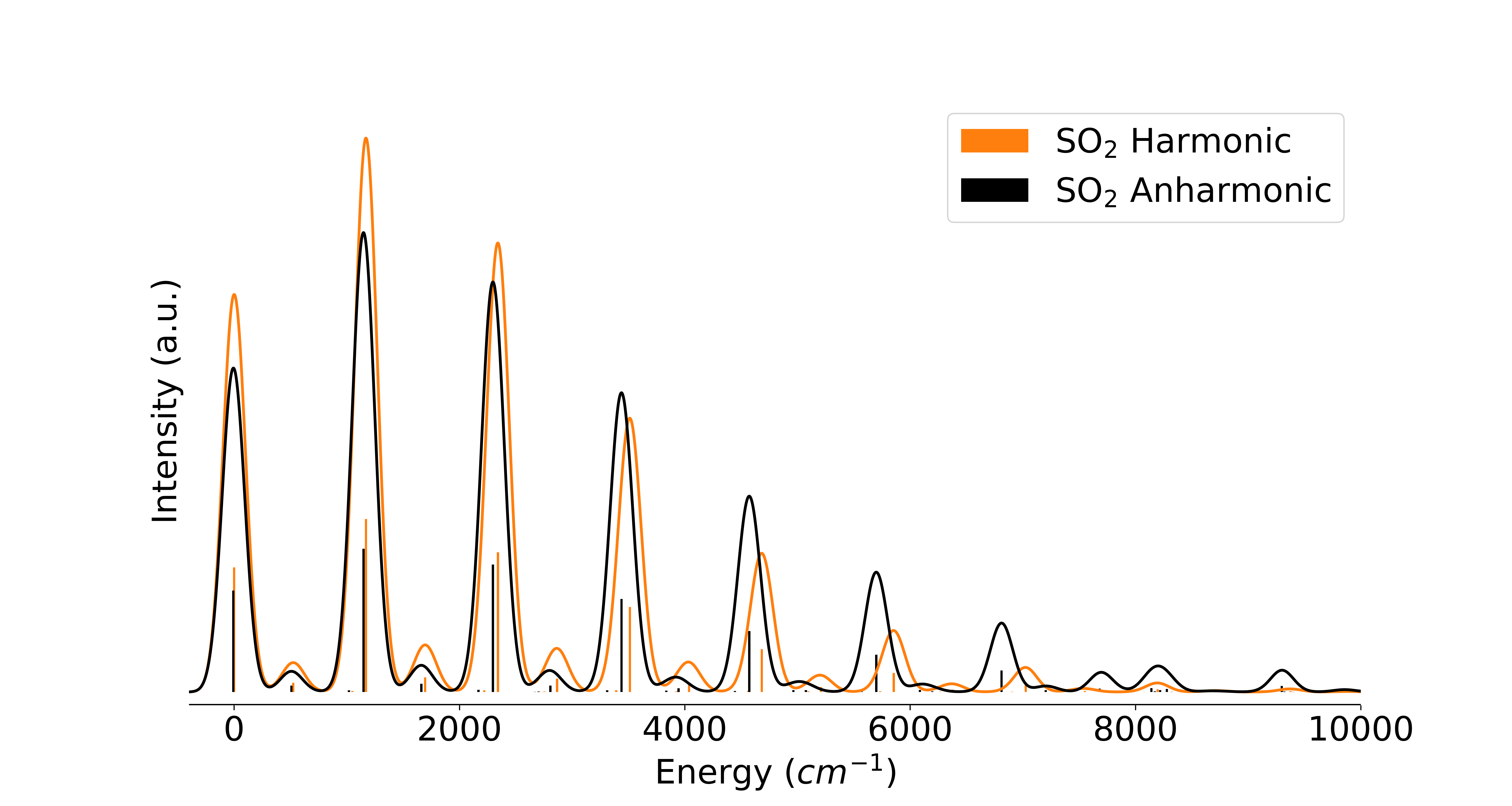}
      \caption{Our quantum algorithm will show the largest performance advantages over classical computers when simulating anharmonic systems. Here we give an example of an anharmonic vibronic simulation (black) for \sox$^{-} \rightarrow$ \sox, showing that the harmonic approximation (orange) is inaccurate for this transition. Vertical lines represent the stick spectra, which are broadened to an arbitrary 100 \invcm. Using the expansion in Eq. \ref{eq:anharm}, the Duschinsky transformation and \sox$^-$~ PES from Lee et al. \cite{lee09}, and the anharmonic \sox~PES from Smith et al. \cite{smith84}, we included the following third- and fourth-order terms: $q_1q_1q_1$, 
$q_1q_1q_2$, 
$q_1q_2q_2$, 
$q_1q_3q_3$, 
$q_2q_2q_2$, 
$q_2q_3q_3$, 
$q_1q_1q_1q_1$, 
$q_1q_1q_2q_2$, 
$q_1q_1q_3q_3$, 
$q_2q_2q_2q_2$, 
$q_2q_2q_3q_3$, and 
$q_3q_3q_3q_3$, where subscripts label the normal mode (see SI for additional details). Many molecules exhibit substantial anharmonicity, which is the class of molecules for which our quantum algorithm will be able calculate spectra that classical algorithms are likely unable to. }
      \label{fig:so2-anharm}
\end{figure}

The least-studied source of error in our algorithm is due to an insufficiently large QHO cutoff \lmax. It is especially important to study this source of error, both qualitatively and quantitatively, because the standard classical algorithms for calculating FCFs  \cite{ruhoff00,jankowiak07,barone09,luis06,huh10,meier15,petrenko17} do not directly simulate the vibrational Hamiltonian in the Fock basis of PES $A$, and hence do not suffer from this type of truncation error. 
An analysis of Suzuki-Trotter errors will be dependent on the QHO mapping chosen and is left to future work.


We chose transitions in four triatomic molecules---sulfur dioxide (\sox$^{-} \rightarrow$ \sox) \cite{lee09}, water (H$_2$O$ \rightarrow $H$_2$O$^+$+e$^-$)\cite{chang08}, deuterated water (\dox; D $\equiv$ $^2$H)\cite{chang08}, and nitrogen dioxide (\nox[$^2A_1$] $\rightarrow  $ \nox[$^2B_2$])\cite{ruhoff94}---and simulated their vibronic spectra using one electronic transition from each (the SI includes additional information, including all physical parameters used). We must note that there is a well-studied conical intersection near the bottom of PES $^2B_2$ in \nox \cite{kirmse98,santoro99,arasaki10,worner11} which greatly alters the vibronic properties---in the approximation used here, this feature is ignored. The latter three molecules were chosen explicitly because they have unusally high phonon occupation numbers for a vibronic transition, making them good candidates for a study on $L_{max}$ requirements.

Fig. \ref{fig:4vibspec} shows both the theoretically exact vibronic spectra (solid line) and an arbitrarily chosen approximate spectrum (dotted line) for each molecule. 
These plots show the qualitative error behavior: higher-energy peaks are blue-shifted while low-energy peaks converge rapidly. It is useful to consider this error trend when implementing the algorithm on a future quantum computer. Simulation details and additional analysis are given in the SI.



Finally, in order to highlight the simulation of anharmonicity as our quantum algorithm's primary application, we consider the zero temperature anharmonic spectrum of \sox$^{-} \rightarrow$ \sox, using the same Duschinsky matrix as before \cite{lee09} but an anharmonic PES for \sox~ taken from Smith et al. \cite{smith84}. Fig. \ref{fig:so2-anharm} shows the stark difference between the full anharmonic simulation and the harmonic approximation. We included both third-order and fourth-order Taylor series terms (Eq. \ref{eq:anharm}), which are easily mapped to the quantum computer using the same procedure as before (see SI for additional details). This relative failure of the harmonic approximation is not uncommon \cite{luis06,jankowiak07,huh10,meier15,petrenko17}, indicating a well-defined set of molecules (those with substantially anharmonic PESs) for which our quantum algorithm would outperform classical computers.

We introduced a quantum algorithm for calculating the vibronic spectrum of a molecule to arbitrary precision. We noted several advantages over the previously proposed vibronic boson sampling (VBS) algorithm. First and perhaps most importantly, anharmonic effects (whose inclusion is very costly classically but often chemically relevant) can be easily included in our approach. Second, measuring the eigenenergy in our algorithm leaves the quantum state preserved, allowing for further analysis that would not be possible in VBS. Third, there are error-related advantages to our approach. 
Aspects of our algorithm may be extended to other chemical processes for which nuclear degrees of freedom are difficult to simulate on a classical computer. This work's general strategy, of calculating the energy distribution outputted from quantum phase estimation to arbitrary precision, may be applied to other spectral problems in chemistry and condensed matter physics.

\section*{Supporting Information.}
Elaboration on vibronic Hamiltonian construction, quantum harmonic oscillator to qubit mappings, the finite temperature algorithm, computational scaling, molecular data and parameters used, and additional error analysis.

\section*{Acknowledgements}
J.H. acknowledges support by the Basic Science Research Program through the National Research Foundation of Korea (NRF) funded by the Ministry of Education, Science and Technology (NRF-2015R1A6A3A04059773). The authors thank Gian Giacomo Guerreschi and Daniel Tabor for helpful suggestions on the manuscript.

\bibliography{main}

\end{document}




\title{\textit{Supporting Information:} Quantum Algorithm for Calculating Molecular Vibronic Spectra}

\author{Nicolas P. D. Sawaya}
\email{nicolas.sawaya@intel.com}
\affiliation{Intel Labs, Santa Clara, California, USA}

\author{Joonsuk Huh}%
 \email{joonsukhuh@gmail.com}
 \affiliation{Department of Chemistry, Sungkyunkwan University, Republic of Korea; \\SKKU Advanced Institute of Nanotechnology (SAINT), Sungkyunkwan University, Republic of Korea}%





\date{\today}


\maketitle

\tableofcontents

\section{Hamiltonian construction} \label{apx:ham}

Here we give a more pedagogical summary of the Hamiltonian construction summarized in the main text. The procedure involves these three transformations: $\{\vec q_A, \vec p_A\}$ $\rightarrow \{\vec Q_A, \vec P_A\}$ $\rightarrow \{\vec Q_B, \vec P_B\}$ $\rightarrow \{\vec q_B, \vec p_B\}$. Mass-weighted position and moment operators, $\vec Q_s$ and $\vec P_s$ respectively, are \cite{huh11thesis}

\begin{equation}
\vec Q_s = \mathbf{\Omega}_s^{-1} \vec{q_s}
\end{equation}

\begin{equation}
\vec P_s = \mathbf{\Omega}_s \vec{p_s}
\end{equation}

with the $M \times M$ matrix

\begin{equation}
\mathbf{\Omega}_s = diag( [\omega_{s1},...,\omega_{sM}] )^{\half}
\end{equation}

where $\{\omega_{sk}\}$ are the scalar harmonic oscillator frequencies of normal mode $k$ on PES $s$. 

Because the Duschinsky transformation is not dimensionless, its direct application is appropriate only to the vector of mass-weighted position and momentum operators:

\begin{equation}
\vec Q_B = \textbf S \vec Q_A + d
\end{equation}

\begin{equation}
\vec P_B = \textbf S \vec P_A.
\end{equation}

Then the following are used to obtain the final dimensionless operators:

\begin{equation}
\vec q_B = \mathbf \Omega_B \vec Q_B
\end{equation}

\begin{equation}
\vec p_B = \mathbf \Omega_B^{-1} \vec P_B.
\end{equation}

Combining these steps leads to equations the following formulas from the main text

\begin{equation}\label{eq:qB}
\vec q_B = \mathbf \Omega_B \mathbf S \mathbf\Omega_A^{-1} \vec q_A + \mathbf \Omega_B \vec d 
\end{equation}

\begin{equation}\label{eq:pB}
\vec p_B = \mathbf \Omega_B^{-1} \mathbf S \mathbf\Omega_A  \vec p_A,
\end{equation}

An alternative route for expressing $H_B$ in terms of the operators of PES $A$ (the one taken in references \cite{malmqvist98,huh15}) first transforms the ladder operators directly using the transformation

\begin{equation} \label{eq:a-j-d}
\vec{b}^\dag = \half (J - (J^t)^{-1})\vec{a} + \half(J+(J^t)^{-1})\vec{a}^\dag + \frac{1}{\sqrt{2}}\vec\delta
\end{equation}

where $\{\hat{a}_i^\dag\}$ and $\{\hat{b}_i^\dag\}$ are respectively creation operators for states of PES $A$ and $B$, and $J = \mathbf{ \Omega_B S \Omega_A^{-1} }$. Eq. \ref{eq:a-j-d} is then used to construct the Hamiltonian $H_B = \sum_i \omega_{Bi} (b_i^\dag b_i + \frac{1}{2})$.


It is important to note that there are oftentimes only one or a few electronic transitions that are relevant for a chemist, often the transition between the ground and first excited state. The potential energy surface (PES) of two electronic states must be calculated beforehand, with one of several electronic structure algorithms. For most organic molecules, density functional theory calculations (which roughly speaking often have cubic scaling in the number of electrons) typically provide electronic PESs that are accurate enough to produce vibronic spectra that match experiment. For other classes of molecules, substantially more expensive methods may be required for obtaining the PES \cite{helgaker14}.

\section{QHO to qubit mappings} \label{sec:qho}

To implement the algorithm within the standard quantum circuit model, one requires a mapping of quantum harmonic oscillators to a set of qubits. Several mappings from bosonic DOFs to qubits have been proposed in the past \cite{somma05,macridin18a,macridin18b,mcardle18}. Here, we outline what are perhaps the two most straightforward mappings for the QHO, which in this work we will call the standard binary and the unary mappings. It is worth mentioning that we would not expect an approach based on the Holstein-Primakoff transformation \cite{HolsteinPrimakoff} to be particularly promising, since it would require first mapping a bosonic system to a spin-$s$ system, after which one would need the additional step of mapping to spin-half qubits using Clebsch-Gordon coefficients.

Here we summarize how one would convert the operators of $H_B$ into quantum gates of the standard circuit model. The mappings are used to represents operators $\tilde a_i^\da$ and $\tilde a_i$ in qubits.

The \textit{standard binary} mapping represents each level as a binary number such that any integer is represented as $\sum_{p=0}^{p_{max}-1} x_p 2^{p}$, where $p$ is the qubit id. The state $[|0\ra$, $|1\ra$, $|2\ra$, $|3\ra$, $|4\ra$, $...]^T$ is isomorphic to $[|000\ra$, $|001\ra$, $|010\ra$, $|011\ra$, $|100\ra$, $...]^T$, where a mapping to 3 qubits was used in this example. Hence each QHO eigenlevel $l$ is a string of $0$s and $1$s. Any single-mode operator used in constructing Hamiltonian $H_B$ can be expressed in terms of elements $| l \ra\la l' |$, where $l$ and $l'$ denote two vibrational levels. In qubit space this leads to operators of the form $|x_0...x_{p_{max}}\ra\la x_0'...x_{p_{max}}'|$ where each $x_p$ is a binary value and $p_{max}$ is the number of qubits used in the mapping for a particular mode. As $|x_0...x_{p_{max}} \ra\la x_0'...x_{p_{max}}'|$ is equivalent to $| x_0 \ra\la x_0' |\otimes\ldots\otimes| x_{p_{max}} \ra\la x_{p_{max}}' |$, in the latter expression one of four operators is substituted for each single-qubit operator:

\begin{equation}\label{eq:op01}
|0\ra\la1| = \half (X + iY) [= \sigma^-] 
\end{equation}

\begin{equation}\label{eq:op10}
|1\ra\la0| = \half (X - iY) [= \sigma^+] 
\end{equation}

\begin{equation}\label{eq:op00}
|0\kb 0| = \half (I + Z)
\end{equation}

\begin{equation}\label{eq:op11}
|1\kb 1| = \half (I - Z)
\end{equation}

where $\{X,Y,Z\}$ are the Pauli operators, and $I$ is the identity operator. In the standard binary mapping, every term $|l'\ra \la l|$ leads to a qubit-space operator that operates on all $p_{max}$ qubits.

The less compact \textit{unary} encoding (for which the earliest reference we are aware of is \cite{somma05}) maps $[|0\ra, |1\ra, |2\ra, |3\ra, |4\ra, ...]^T$ to $[|00001\ra, |00010\ra, |00100\ra, |01000\ra, ...]$, requiring more qubits but fewer gates to implement an operator. There are $L_{max}+1$ qubits in this mapping, as one qubit is reserved for the vacuum state. Though the standard binary mapping utilizes the full Hilbert space, the unary code uses only a small subspace of it. As a result, individual terms of the number operator, i.e. $l|l \ra\la l|$, are represented by a single qubit operator using Eq. \ref{eq:op11}; nearest-level terms like $|l+1\ra\la l|$ can be represented by two-qubit operators $\sigma^-_l\sigma^+_{l+1}$.

In real-world implementations, the choice of mapping is likely to depend on a given hardware's qubit count and connectivity. In near-term devices without error correction, the coherence time will have to be considered as well, as different mappings produce circuits of differing depths. Detailed analysis of the cost of each mapping, for a given $L_{max}$, is deferred to future work, as this requires detailed consideration of circuit optimization, gate cancellations, and qubit connectivity constraints.



The quantum circuit model requires us to set a finite cutoff for the maximum occupation number of each QHO. For vibronic transitions in real molecules, the number $l_j$ of vibrational quanta in the $j$th mode does not exceed some maximum value $L_{max,j}$ (assuming some finite precision) \cite{jankowiak07}. In practice, on a future real-world quantum computer, the simplest solution is to increase $L_{max,j}$ values for all modes until convergence is reached.




\section{Finite Temperature Algorithm}\label{sec:apxFT}
Finite temperature effects can be included by appending additional steps before and after the zero temperature algorithm, in line with previous work \cite{huh17ft}. 
The idea is to begin with a purification of the mixed state of the Bolzmann distribution, by having each independent mode be represented by two subspaces in a purified Fock state. 
It is necessary to introduce the scalar function $E_A(\nb)$, defined as the energy of a Fock state in PES $A$:
\begin{equation}\label{eq:EA}
    E_A(\nb) = E_A([n_0,...,n_M]) = \sum_i \omega_i (n_i + \half),
\end{equation}
where $n_i$ is the occupation number of the $i$th mode.

First we add an additional register, $I$ (for `initial state'), of ancilla qubits. Registers $I$ and $S$ must have the same size, and we prepare a pure state $|\Psi_{IS}\ra = \sum|\phi\ra_I|\psi\ra_S$ such that $\rho_{th} = Tr_I(|\Psi_{IS}\ra\la\Psi_{IS}|)$ is the desired Gibbs thermal state in the initial PES. Before running the QPE routine, we need the pure state
\begin{equation}\label{Psi_IS}
|\Psi_{IS}\ra 
= \sum_{\nb} \kappa_\nb|\nb\ra_I \otimes |\nb\ra_S
= \sum_{\nb} \sqrt{\la \nb |\rho_{th}| \nb \ra}|\nb\ra_I \otimes |\nb\ra_S 
\end{equation}

where $|\nb\ra = |n_0,...,n_M\ra$. To prepare $|\Psi_{IS}\ra = \hat V(\beta) |\mathbf{0}\ra_I|\mathbf{0}\ra_S$, one implements the unitary operator
\begin{equation}\label{eq:vbeta}
    \hat V(\beta) = \bigotimes_i^M \exp( \theta_i (\alpha_i^\da a_i^\da - \alpha_i a_i)/2)
\end{equation}

where $\alpha_i^\da$ and $\alpha_i$ are ladder operators for the $i$th vibrational mode of register $I$. The inverse temperature is $\beta=1/k_B T$, where $k_B$ is the Boltzman constant and $T$ is temperature. Angle $\theta_i$ is defined by $\tanh(\theta_i/2)$ $=\exp(-\beta\hbar\omega_i/2)=\sqrt{\overline{n}_i/(\overline{n}_i+1)}$ and $\overline{n}_i$ is the mean quantum number for mode $i$ \cite{mann89,huh17ft}. This operator can be applied using the previously discussed procedure to map arbitrary bosonic operators to qubit operators (Section \ref{sec:qho}).

After this initial state preparation step, the remainder of the algorithm proceeds as before, but with the following additional elements. After the QPE circuit is applied using registers $S$ and $E$ as before, registers $I$ and $E$ are both measured. The measured state $|\nb_I\ra$ in register $I$ effectively acts as a label, indicating the vibrational eigenstate (Fock state) in the initial PES from which the measured transition occurred. Finally, the contribution to the vibronic spectrum is $\tildeps_i - E_A(\nb_I)$, instead of just $\tildeps_i$, because the measured transition ``began'' in vibrational state $|\nb_I\ra$ in the $A$ basis. An outline of the procedure is given in Appendix \ref{sec:algoutlines} and a quantum circuit diagram is shown in Fig. 2 (main text). For anharmonic PESs, a similar procedure would be used, with an appropriate anharmonic preparation step used in place of Eq. \ref{eq:vbeta} \cite{christiansen04,poulin09}.

\section{Outline of Algorithms} \label{sec:algoutlines}

What follows is an outline of the zero- and finite-temperature algorithms for calculating molecular vibronic spectra.

Some conceptual clarifications are worth noting. In both the zero- and finite-temperature algorithms, the procedure is to produce a histogram with an arbitrary energy resolution, determined by the number of bits used in quantum register $E$. Quantum superposition is the key to the algorithm; it removes the need to consider each state explicitly. Even in the finite temperature case, one does not explicitly consider every non-negligible state of the Boltzmann distribution---one prepares a superposition all the possibilities for initial and final states, and then samples their energies. The problem is effectively reduced to sampling from a one-dimensional probability distribution corresponding to the (zero- or finite-temperature) vibronic energy spectrum.

\textbf{Zero-temperature algorithm:}
\begin{enumerate}
\item Initialize state $|0\ra_S|0\ra_E$.
\item Run QPE using Hamiltonian $H_B$, expressed in the $A$ basis: $|0\ra|0\ra \rightarrow \sum_i c_i |\psi_i\ra|\tildeps_i\ra$.
\item Measure register $E$ to obtain eigenenergy $\tildeps_i$: $\sum_i c_i|\psi_i\ra|\tildeps_i\ra \rightarrow A_j\sum_{k \in \mathcal{D}_j}c_k|\psi_k\ra|\tildeps_j\ra$, where $A_j$ is a renormalization constant.
\item \label{itm:analys} If desired, perform additional analysis on the preserved state $A_j\sum_{k \in \mathcal{D}_j}c_k|\psi_k\ra$ in register $S$, as discussed in the main text. For example, perform a SWAP test with another state, or resolve one of the Fock states in $\mathcal D_j$.
\item Repeat these steps to obtain a histogram of $\tildeps_i$ values.
\end{enumerate}

\textbf{Finite-temperature algorithm:}
\begin{enumerate}
    \item Initialize state $|0\ra_I|0\ra_S|0\ra_E$.
    \item Prepare the thermal state by acting on registers $I$ and $S$: $\hat V(\beta)|0\ra|0\ra|0\ra \rightarrow \sum_\nb \kappa_\nb|\nb\ra|\nb\ra|0\ra$.
    \item Apply QPE with $H_B$, on registers $S$ and $E$: $\sum_\nb \kappa_\nb|\nb\ra|\nb\ra|0\ra $  $\rightarrow$ $\sum_\nb \kappa_\nb|\nb\ra  \sum_i (c_{\nb,i} |\psi_i\ra |\tildeps_i\ra)$.
    \item Measure both registers $E$ and $I$: $\rightarrow |\nb_I\ra (A_{j,\nb_I}\sum_{k \in \mathcal{D}_j}c_{\nb_I, k}|\psi_k\ra) |\tildeps_j\ra$.
    \item Perform optional analysis on register $S$, as previously mentioned.
    \item The contribution to the histogram is then $\tildeps_j - E_A(\nb_I)$. (Contrast this with the zero-temperature case, where $E_A(\nb_I)$ is omitted.) 
\end{enumerate}

\section{Computational Scaling}\label{sec:scaling}

Below we assume the parameters $\mathbf S$, $\vec\delta$, $\mathbf \Omega_A$, and $\mathbf \Omega_B$ are given. Setting aside more advanced linear algebra approaches, both the $q$-$p$ construction ($H_B = \frac{1}{2} \sum_k^M \omega_{Bk} ( q_{Bk}^2 + p_{Bk}^2)$) and the ladder operator construction ($H_B = \sum_i \omega_{Bi} [b_i^\dag b_i + \frac{1}{2}]$) require $O(M^2)$ classical preparation steps, since all transformations involve only matrix-vector multiplications or diagonal-dense matrix multiplications. For comparison, VBS requires $O(M^3)$ classical steps because it uses the singular value decomposition. 
As described in the main text, one element of our algorithm uses Hamiltonian simulation to implement $H_B$ for use in the quantum phase estimation (QPE) algorithm.  An essential consideration, especially for near- and mid-term hardware, is the computational cost of implementing one Trotter step of the Hamiltonian's propagator.  


Each operator $b_i$ is a linear combination of terms in $\{a_i^\dag\}$ and $\{a_i\}$. The operator $H_{B}$, after summing the number operators in $\{b_i^\dag b_i\}$ and grouping terms, is a linear combination of terms in $\{a_i a_j\}$,$\{a_i^\dag a_j\}$, $\{a_i a_j^\dag\}$, and $\{a_i^\dag a_j^\dag\}$. Hence in the worst case, the number of terms in $H_{B}$ scales as $O(M^2)$, meaning the number of operations in a Trotter step is $O(M^2)$ as well.

The circuit depth of a Trotter step scales as $O(M)$, \textit{i.e.} linear-depth. To see this, consider placing two-boson operators (each corresponding to an interaction term such as $a_i^\dag a_j$) on all boson pairs ${i,j}$ that satisfy $(i-j)=w\mod N_q$, where $w \in \{1,2,...,N_q-1\}$. For a single value of $w$, this gate placement has constant depth. Iterating through all values of $w$ yields a circuit with linear depth $O(M)$, and single-boson terms do not change this scaling. The same argument applies to the method based on $\hat q$ and $\hat p$ operators. Note that the finite-temperature algorithm scales no worse than the zero-temperature procedure, since the state preparation takes $O(M)$ operations with $O(1)$ depth.

When anharmonic effects are included, the complexity of implementing a Trotter step will be $O(M^k)$, where $k$ is the highest-order term in the Taylor expansion of the anharmonic Hamiltonian. It is possible that there will be methods for reducing this complexity in the anharmonic case, for example by using other other classes of functions in the expansion, \textit{e.g.} the Morse potential.

\section{Molecular data}\label{sec:molecdata}
The four simulated molecules, all of the C$_{2v}$ point group, have three vibrational modes: a bending mode, a symmetric stretch, and an anti-symmetric stretch. Due to symmetry, the first two modes are decopuled from the anti-symmetric mode, assuming the harmonic approximation. We consider only the two coupled modes in the harmonic analyses.

For all molecules other than \nox, we are effectively calculating the photoelectron spectrum, as we are considering an ionization process.  Additionally, because of the experimental difficulty in photon counting for higher occupation numbers, in the future it is possible that these molecules might be more easily simulated on a universal quantum computer than a photonic device \cite{walmsley17,shen18}. The electronic transitions are SO$_2^- \rightarrow $SO$_2+$e$^-$ \cite{lee09}, H$_2$O(D$_2$O)$ \rightarrow $H$_2$O$^+$(D$_2$O$^+$)+e$^-$ \cite{chang08}, and \nox's ground to excited state transition $^2A_1 \rightarrow  $ $^2B_2$ \cite{ruhoff94}.

The following parameters were used, taken from the literature. $\mathbf S$ and $\delta$ are dimensionless; energies of $\vec{\omega}$ are in wavenumbers, $cm^{-1}$.

$SO_2^- \rightarrow SO_2+e^-$ \cite{lee09}: 
\begin{equation} \label{eq:smat_so2}
\mathbf S_{SO_2} = 
\begin{bmatrix}
0.9979  & 0.0646 \\
-0.0646 & 0.9979 \\
\end{bmatrix}
\end{equation}

$$
\delta_{SO_2} = 
\begin{bmatrix}
-1.8830 \\
0.4551 \\
\end{bmatrix} 
$$
$$
\vec{\omega}_{SO_2^-} =
\begin{bmatrix}
943.3 \\
464.7
\end{bmatrix}
$$
$$
\vec{\omega}_{SO_2} =
\begin{bmatrix}
1178.1 \\
518.8
\end{bmatrix}
$$

$H_2O \rightarrow H_2O^+ +e^-$ \cite{chang08}:
\begin{equation} \label{eq:smat_h2o}
\mathbf S_{H_2O} = 
\begin{bmatrix}
0.9884 & -0.1523 \\
0.1523 & 0.9884
\end{bmatrix}  
\end{equation}

$$
\delta_{H_2O} = 
\begin{bmatrix}
0.5453  \\
4.2388  \\
\end{bmatrix}  
$$
$$
\vec{\omega}_{H_2O} =
\begin{bmatrix}
3862 \\
1649
\end{bmatrix}  
$$
$$
\vec{\omega}_{H_2O^+} =
\begin{bmatrix}
2633 \\
1620
\end{bmatrix}
$$

$D_2O \rightarrow D_2O^+ +e^-$ \cite{chang08}:
\begin{equation} \label{eq:smat_h2o}
\mathbf S_{D_2O} = 
\begin{bmatrix}
0.9848 & -0.1737 \\
0.1737 & 0.9848
\end{bmatrix}  
\end{equation}

$$
\delta_{D_2O} = 
\begin{bmatrix}
0.7175 \\
4.8987 \\
\end{bmatrix}  
$$
$$
\vec{\omega}_{D_2O} =
\begin{bmatrix}
2785 \\
1207
\end{bmatrix}  
$$
$$
\vec{\omega}_{D_2O^+} =
\begin{bmatrix}
1915 \\
1175
\end{bmatrix}
$$

\nox ($^2A_1 \rightarrow $ $^2B_2$) \cite{ruhoff94}:
\begin{equation} \label{eq:smat_no2}
\mathbf S_{NO_2} = 
\begin{bmatrix}
0.938 & -0.346 \\
0.346 & 0.938  \\
\end{bmatrix}  
\end{equation}
$$
\delta_{NO_2} = 
\begin{bmatrix}
-4.0419 \\
 5.3185  \\
\end{bmatrix}  
$$
$$
\vec{\omega}_{NO_2(gr)} =
\begin{bmatrix}
1358 \\
757
\end{bmatrix}  
$$
$$
\vec{\omega}_{NO_2(ex)} =
\begin{bmatrix}
1461 \\
739
\end{bmatrix}
$$


For our anharmonic simulation, we used the same Duschinsky matrix as before \cite{lee09}, but used the anharmonic PES for the electrically neutral \sox~ from Smith et al. \cite{smith84}. We use a harmonic potential energy surface only for the initial PES \sox$^-$, which is a good approximation because the initial vibrational state is in the ground state (Fock vacuum state) of \sox$^-$, \textit{i.e.} the initial state is at the bottom of the PES, where the harmonic approximation is valid. Additionally, the third vibrational mode can no longer be considered decoupled when anharmonic effects are included, making this a simulation of all three vibrational modes. Hence, for the anharmonic spectrum, the following parameters are taken from Lee et al. \cite{lee09}:

\begin{equation} \label{eq:smat_so2_anharm}
\mathbf S_{SO_2^- \rightarrow SO_2} = 
\begin{bmatrix}
0.9979  & 0.0646 & 0 \\
-0.0646 & 0.9979 & 0 \\
   0   &    0   & 1 \\
\end{bmatrix}
\end{equation}

\begin{equation}
\delta_{SO_2^- \rightarrow SO_2} = 
\begin{bmatrix}
-1.8830 \\
0.4551 \\
0 \\
\end{bmatrix} 
\end{equation}

\begin{equation}
\vec{\omega}_{SO_2^-} =
\begin{bmatrix}
943.3 \\
464.7 \\
1138.6
\end{bmatrix}
\end{equation}

And from Smith et al. \cite{smith84}:
\begin{equation}
\vec{\omega}_{SO_2}^{anharm} =
\begin{bmatrix}
1171 \\
525 \\
1378
\end{bmatrix}
\end{equation}

We then include the third- and fourth-order terms in the Taylor expansion (Eq. 8 in the main text). Table \ref{tab:so2_anharm} gives the coefficients for the anharmonic terms $q_1q_1q_1$, 
$q_1q_1q_2$, 
$q_1q_2q_2$, 
$q_1q_3q_3$, 
$q_2q_2q_2$, 
$q_2q_3q_3$, 
$q_1q_1q_1q_1$, 
$q_1q_1q_2q_2$, 
$q_1q_1q_3q_3$, 
$q_2q_2q_2q_2$, 
$q_2q_2q_3q_3$, and 
$q_3q_3q_3q_3$. All of these operators may be mapped to qubit-based Pauli operators using exactly the same procedure that was outlined before (Section \ref{sec:qho}).

\begin{table}[]
    \centering
    \begin{tabular}{| c | c | c | c | c | c | c | c | c | c | c | c |}
       $k_{111}$ & $k_{112}$  &  $k_{122}$  &  $k_{133}$  & $k_{222}$  &   $k_{233}$  &  $k_{1111}$  &   $k_{1122}$  &   $k_{1133}$  &   $k_{2222}$  &       $k_{2233}$  &  $k_{3333}$   \\
       44 & -19 & -12 & 159 & -7.0 & 4.7 & 1.8 & -3.1 & 15 & -1.4 & -6.5 & 3.0 
       
       \end{tabular}
    \caption{Higher-order terms used in the \textit{anharmonic} potential energy surface of the neutral SO$_2$ molecule \cite{smith84}. All values are in units of $cm^{-1}$.}
    \label{tab:so2_anharm}
\end{table}






\section{Error Analysis}\label{sec:apxerr}

\begin{figure}
  \centering
    \includegraphics[width=0.8\textwidth]{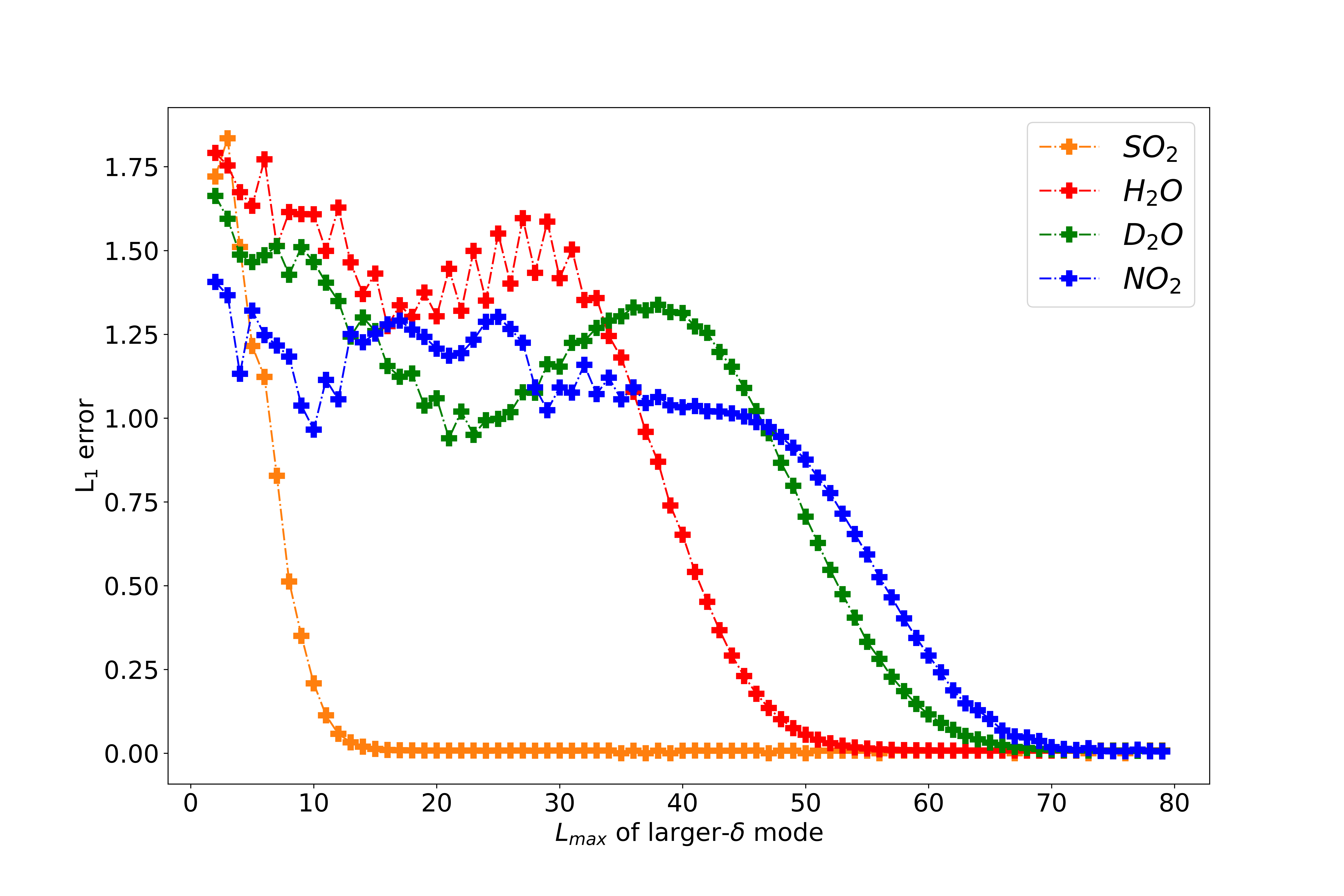}
      \caption{L$_1$-norm errors between exact and approximate vibronic spectra, for molecules $SO_2$, $H_2O$, $D_2O$, and $NO_2$ (where $D$ is deuterium), where each eigenvalue was broadened with a Gaussian of width 100 \invcm~ to make error analysis possible (broadening is performed after the histogram is constructed). $H_2O$, $D_2O$, and $NO_2$ were chosen because they have particularly high phonon occupation numbers, necessitating a large QHO cutoff \lmax. In general a larger displacement $\delta$ leads to a larger required cutoff. In this simulation, the mode with a smaller $\delta$ was assigned a converged \lmax; hence we isolated the effects of the variable of the more significant (larger $\delta$) mode by varying its \lmax. }
      \label{fig:plotL1}
\end{figure}

We studied truncation errors, i.e. those due to insufficiently large \lmax, primarily because this type of error is not present in standard classical vibronic simulations, which are not based on diagonalizing $H_B$ \cite{ruhoff00,jankowiak07,barone09,huh11thesis}. All results are obtained by creating $H_B$ with truncated ladder operators, diagonalizing the Hamiltonian, calculating FCFs, and binning the results in bins of width 1 cm$^{-1}$.

To make our error analysis method possible, the spectra in this work were broadened with a Gaussian of width 100 $cm^{-1}$, a width that represents $\lesssim$ 1\% of the spectral range for these four molecules. The broadening is a distinct separate step, and is performed after formation of the histogram. Errors were calculated using the L$_1$ norm between the exact and approximate spectra (both broadened),

\begin{equation}\label{eq:l1}
\epsilon_{L1} = \int |FCP_{exact}(\omega)-FCP_{approx}(\omega)| d\omega.
\end{equation}
Because FCF profiles have unit norm, the worst case of two spectra with zero overlap yields $\epsilon_{L1} = 2$.


The exact and approximate Hamiltonians were constructed using equation \ref{eq:a-j-d}, varying ladder operator size to reflect \lmax. The numerically exact results were considered converged when the L$_1$ norm between two subsequent $L_{max}$ values was below $10^{-4}$. We validated our method's numerically exact results by demonstrating that our results for \sox ~were identical to those produced by the software program hotFCHT \cite{berger98}, which uses an entirely different algorithmic approach based on recurrence formulas.

For all simulations, the mode that required a smaller cutoff was set to a high converged value, so that we isolated the effect of \lmax ~for the mode requiring a larger cutoff. This is the mode that is more shifted, i.e. the one with larger $|\delta|$. Hence for \sox ~we varied the cutoff for the first mode, while for the other three molecules we varied the cutoff for the second mode. We plotted the approximate spectra (dotted lines, Fig. 3) in order to demonstrate the qualitative effect of an insufficient cutoff. The approximate spectra in Fig. 3 were arbitrarily chosen such that $\epsilon_{L1}$ lies between 0.2 and 0.25. For these illustrative approximate spectra, $\epsilon_{L1}$ and \lmax ~are \{0.208, 0.231, 0.228, 0.241\} and \{10, 45, 57, 61\} for \sox, \hox, \dox, and \nox, respectively.

Qualitatively, the effect of a too-low cutoff number is to preferentially blue shift the higher energy peaks (Fig. 3 in main text). This numerical artifact results from the fact that the $L_{max}$ cutoff effectively introduces anharmonicity to the problem; operators constructed from exact (infinite) ladder operators will not have the same spectrum as those constructed from truncated operators. As \lmax ~is increased, the low energy peaks are converged much sooner than the high energy peaks are. For instance, in the approximate \hox ~spectrum shown, there is an effectively perfect match below $\sim$15,000 $cm^{-1}$, but the blue-shift errors become even larger than $\sim $100 $cm^{-1}$ for eigenvalues above $\sim$23,000 $cm^{-1}$. Being aware of this consistent qualitative error behavior can provide guidance when interpreting results from an implementation of our quantum algorithm. Additional results on convergence with increasing \lmax ~are shown in Section \ref{sec:apxerr}. When using a future quantum computer, one would need to run the algorithm with increasing \lmax ~until the spectrum is converged.

Fig. \ref{fig:plotL1} shows $\epsilon_{L1}$ as a function of \lmax, again for the mode with larger $\delta$. The approximate \lmax~ cutoffs at which the error can be considered converged are [12, 51, 64, 69] respectively for \sox, \hox, \dox, and \nox. For this small set, the \lmax~order matches the order of increasing $\delta$, which is the expected approximate trend. Using the standard binary mapping for QHO levels (which requires $\lceil \log_2 L_{max} \rceil$ for a given mode) would mean that the number of qubits required for the larger-$\delta$ mode are 4, 6, 6, and 7 qubits, respectively. 




Counter-intuitively, $L_{max}$ must be substantially larger than the highest QHO level at which appreciable intensity exists. For example, one may naively expect that $L_{max}$=8 would be sufficient for \sox, since the FC factor $\sum_{n_1'}|\la\mathbf{0}|n_0'=8\ra|^2$ is a near-negligible value of $\sim$ $1.6\times10^{-3}$ (just 0.6\% of the largest FCF). But $L_{max}$=13 is required for eigenvalue positions and the L$_1$-norm error to converge. This is despite the fact that transitions to levels 12 and 13 are very small, with $\sum_{n_1'}|\la\mathbf{0}|n_0'=12\ra|^2\approx 5.2\times10^{-5}$ and $\sum_{n_1'}|\la\mathbf{0}|n_0'=13\ra|^2\approx 1.5\times10^{-5}$.

 The truncation values are not expected to depend explicitly on $M$ because the intensities of a given mode's vibronic progression is known to approximately follow the rapidly-decaying Poisson distribution \cite{may08}.

\bibliographystyle{alpha}
\bibliography{supp}
